\documentclass{ws-rv975x65}
\usepackage{subfigure}     % required only when side-by-side / subfigures are used
\usepackage{ws-rv-van}     % numbered citation/references (default)
\usepackage{epic}
\makeindex
\newcommand{\del}{\partial}

\begin{document}

\makeatletter
\renewcommand\thesection{\arabic{section}}
\renewcommand\thefigure{\@arabic\c@figure}
\renewcommand\thetable{\@arabic\c@table}
\renewcommand\theequation{\@arabic\c@equation}
\makeatother
\chapter*{Quarks and Anomalies}

\author[R.~J.~Crewther]{R.~J.~Crewther}
%\index[aindx]{Author, F.} % or \aindx{Author, F.}
%\index[aindx]{Author, S.} % or \aindx{Author, S.}

\address{Department of Physics, University of Adelaide\\
Adelaide SA 5005, Australia, \\
rodney.crewther@adelaide.edu.au}

\begin{abstract}
A nonperturbative understanding of neutral pion decay was an essential
step towards the idea that strong interactions are governed by a
color gauge theory for quarks. Some aspects of this work and related problems
are still important.
%The abstract should summarize the context, content and conclusions of
%the paper in less than 200 words. It should not contain any references
%or displayed equations. Typeset the abstract in 9 pt Times roman with
%baselineskip of 11 pt, making an indentation of 1.5 pica on the left
%and right margins.
\end{abstract}
%\markright{Customized Running Head for Odd Page} % default is Chapter Title.
\body

\section{Quarks before QCD}\label{ra_sec1}
%
%The WSPC class file has already loaded the packages
%\verb|amsfonts, amsmath,| \verb|amssymb, graphicx, rotating,| and \verb|url|
%at the startup.
%
\setlength{\unitlength}{1mm}
\begin{picture}(0,0)(-100,-100)
ADP--14--39/T898
\end{picture}% 
Any Caltech theory student in the late 1960's, particularly if Murray Gell-Mann 
was their supervisor, had to be good at distinguishing various ``quark models''.
Were we talking about ``constituent'' or ``current'' quarks, and within those
categories, was model dependence an issue? Quarks were somehow fundamental, but
it was not even clear that their dynamics should be governed by a local field
theory. The main tactic was to ``abstract'' rules which seemed to be model 
independent and led to physical consequences which could be compared with 
existing data.

By that time, the quark idea was several years old, dating from work completed 
independently by the end of 1963:  Gell-Mann's quarks,\cite{quarks} Zweig's 
aces,\cite{aces} and (a reference I have just heard of) Petermann's ``spineurs 
(avec) \ldots des valeurs non enti\`{e}res de la charge''.\cite{swiss}  These 
papers had in common 
\begin{enumerate} 
\item structures $q\bar{q}$ for mesons and $qqq$ for baryons built from
non-relativistic \emph{constituent} quarks $q$ and anti-quarks $\bar{q}$,
\label{structure}
\item the idea that $SU(3)$ mass formulas \cite{eightfold,okubo} are due to 
the strange quark $s$ being heavier than the up and down quarks $u,d$, and 
\item concerns about whether the fractional charges would be observable. 
\end{enumerate}
Gell-Mann and Zweig were led to (\ref{structure}) by the need to explain 
the absence of exotic $SU(3)$ multiplets in the Eightfold Way 
\cite{eightfold,yuval}. Zweig analyzed the constituent quark model in
detail, deriving properties such as spins, parities and masses for various 
$SU(3)$ multiplets. Gell-Mann had a separate aim: to reproduce current 
algebra, a set of equal-time commutators for $SU(3) \times SU(3)$ 
currents\cite{algebra} which he had previously managed to abstract 
without using quarks. For this, he needed \emph{current quarks}, i.e.\ 
relativistic fields $q(x)$ and $\bar{q}(x)$ for each flavor $q = u,d,s$, 
from which electromagnetic, weak and other $SU(3) \times SU(3)$ 
currents could be constructed.

Immediately, there were concerns about constituent quark statistics. How can 
a baryon like $\Sigma^{++}$ exist as an $S$-wave spin-flavor symmetric state
$|u$\mbox{$\uparrow$}$u$\mbox{$\uparrow$}$u$\mbox{$\uparrow$}$\rangle$ if 
quarks are spin-$\frac{1}{2}$ fermions?  It is hard to imagine ground states 
being $P$-wave, so instead, it was proposed that quarks are either\cite{wally} 
para-fermions\cite{bert} of order 3 or\cite{boris,han,miya} fermions with an
extra quantum number taking three values, which we now know as 
\emph{color}.\cite{licht,schladming,bfg-m}. The observed fermionic baryons 
$|qqq\rangle$ are then symmetric in space-spin-flavor (a) for paraquarks
automatically, or (b) for fermion quarks antisymmetrized in a color $SU(3)$ 
singlet state\cite{han,miya} (but not $SO(3)$, because that would allow 
colorless diquark states $|qq\rangle$). 

Whether para-particle or colored multiplets would appear at higher energies or
be banned completely (quark confinement) was not clear. In an attempt to make 
these extra states appear less weird, colored quarks were initially given 
integer charges\cite{han,miya} which, however, depended on the color index.
Then photons could excite color from hadrons and perhaps induce transitions
to a deconfined (weird) sector. 

In the model eventually adopted in 1972\cite{schladming,bfg-m}, quarks became
colored fermions with fractional charges, with 3 colors for each charge or
flavor. As a result, the electromagnetic and weak currents became color $SU(3)$ 
singlets, like the observed hadronic spectrum. Confinement was as unclear for 
this model as the others. If confinement were not absolute, the model could 
have degenerate color multiplets and fractionally charged states above some 
threshold energy. Comparing all of these models, it was concluded that, as 
models of \emph{constituent} quarks, they were hard to distinguish below 
thresholds for deconfinement.

However the 1972 model was also designed to take into account color for 
\emph{current} quarks. The rest of this article describes how studies of 
short-distance behavior\cite{kgw} and the reaction\cite{rjc} 
$\pi^0 \to \gamma\gamma$ led to this.

\section{Scale Invariance at High Energies}

I started life at Caltech as a graduate student in the fall of 1968. The
very first seminar, on Tuesday October 1, was ``Partons'' by Richard P.\
Feynman, with Murray Gell-Mann sitting near the front. Feynman had just
returned from a summer in SLAC hearing about Bjorken's work\cite{bj} on
scaling in deep inelastic lepton-nucleon scattering and developing a 
model of point scatterers (partons) to give the same results. Murray 
kept asking ``but Richard, what are their quantum numbers? Are they quarks?'' 
but Richard's sole concern was scaling due to scattering by ``grains 
of sand inside the nucleon''. (A year or so later, my fellow student 
Finn Ravndal got him interested in quarks.) 

Murray began supervising me two months later and in due course asked me, 
as an initial research exercise, to try using the Cutkosky bootstrap model
\cite{cut} to generate higher symmetries like $SU(6)$. That produced 
hundreds of equations. Fortunately, just a few of them could be used to 
show that there could be no consistent solution. Murray commented that 
he hadn't intended the exercise ``to be so vigorous'' and suggested 
that I take a trip while he thought of a suitable PhD topic. My fellow 
student Chris Hamer and I had already planned to drive around the US 
that summer (1969), so we left immediately and on the way back, stopped 
at Aspen.   

Murray had just started working on scale invariance as an approximate
symmetry of hadrons, and suggested that I do the same. This would involve
the energy-momentum tensor $\theta_{\mu\nu}$ as well as the $SU(3) \times
SU(3)$ currents. Did I know about the Belinfante\cite{bel} tensor? 
Fortunately, I did (from Geoff Opat, supervisor of Chris and myself as
Masters students in Melbourne, 1966-68). In that case, the next step was
to understand all 14 pages of Wilson's paper on operator product 
expansions\cite{kgw}.

Wilson generalised current algebra, replacing equal-time limits of
commutators by short-distance limits of products of currents and
other observables such as $\theta_{\mu\nu}$. Instead of a single
term on the right-hand side, he obtained an asymptotic expansion 
$\sum_n\mathcal{C}_nO_n$ with coefficient functions 
\begin{equation}
\mathcal{C}_1 \gg \mathcal{C}_2 \gg \mathcal{C}_3 \gg\ \ldots
\label{coeff}
\end{equation}
in order of decreasing singularity times observable operators 
$O_1, O_2, O_3 \ldots$ of increasing operator dimensionality (in mass
units). Equal-time commutators, such as in Gell-Mann's current algebra 
and Bjorken's work on scaling, could be recovered by noting that, since 
commutators vanish for space-like separations, their equal-time
limits are controlled by the short-distance behavior of the relevant
operator product. Checks in renormalized perturbation theories or for
free current quarks indicated that, apart from quantum number 
constraints, the same set of operators $\{O_n\}$ tended to appear
in the expansion, whatever the operator product used to generate them:
``a limited set of licensed operators'', as Murray put it.

A key feature of Wilson's work was his critique\cite{kgw2} of canonical 
field theory: operators usually \emph{cannot} be multiplied at the same 
point, equal-time commutators may be singular, and $T$-ordering with
step functions $\theta(t - t')$ can fail. These \emph{anomalies} arise
wherever renormalization is necessary. In particular, renormalized
perturbation theory produces  $\log^p\bigl(\mu^2(x-y)^2\bigr)$ 
factors at short distances, where $\mu$ is the renormalization scale.
When summed up \`{a} la Gell-Mann and Low,\cite{gml} anomalous powers
may be produced. If the ultraviolet limit is controlled by a
nontrivial Gell-Mann--Low fixed point, scale invariance becomes exact 
at short distances, with anomalous dimensions for all operators $O_n$ 
except those which are conserved or partially conserved. I was happy
to abstract these rules and learn the renormalization group later.

Wilson's paper\cite{kgw} also featured a Sec.\ VII ``Applications'' 
with five subsections, each equivalent to a separate publication. 
Subsection D ``$\pi^0 \to \gamma\gamma$ Problem'' drew my attention 
because (a) it explained how short-distance singularities
determine contact terms in low-energy Ward identities and (b) I had
seen the papers of Bell and Jackiw\cite{bell} and Adler\cite{steve} 
on the axial anomaly. Could the three-point function 
$T\langle\mbox{vac}|J_\alpha J_\beta J_{\mu 5}|\mbox{vac}\rangle$
of the electromagnetic and axial-vector currents $J_\alpha$ and $J_{\mu 5}$
be determined at short distances without using perturbation theory? 
Noting Wilson's comment (Sec.\ VIII) that ``the prospects for obtaining 
such a solution seem dim at present'', I filed the problem away as 
a challenge for the future.  

At that time, the main question was whether Bjorken scaling is exact or 
not. Bjorken\cite{bj} obtained scaling by assuming that an infinite set of 
equal-time commutators of $J_\alpha$ with its derivatives is finite, i.e.\
not zero. It was quickly established that this was equivalent to assuming
canonical or free-field (parton) behavior for the coefficient functions 
(\ref{coeff}). Towers of these short-distance singularities could be 
summed to form terms in an operator product expansion near the light 
cone $(x - y)^2 \to 0$, the limit in position space conjugate to Bjorken's 
limit.\cite{frish,bp}  By then, quarks were widely believed to be 
responsible for scaling, so the proposal of Fritzsch and Gell-Mann\cite{fgm} 
to abstract the light-cone expansion from free-quark theory was logical.

The argument against exact Bjorken scaling was led by Wilson.\cite{kgw3}
Interactions tend to increase the dimensions of composite-field operators
$O_n$ which are not conserved exactly or partially, making higher-$n$
functions (\ref{coeff}) less singular on the light cone. The difficulty 
for this point of view was explaining why these anomalous corrections 
were \emph{all} so small. Nevertheless, I tended to belong to this school
of thought. My concern was that any tensor operator $O(x)$ lacking an anomalous 
dimension would be at least partially conserved, because the leading 
singularity of $\langle\mbox{vac}|O(x)O(y)|\mbox{vac}\rangle$
would be canonical and hence divergenceless. Therefore, my view was
that the \emph{only} operators allowed to have canonical dimension were
$\theta_{\mu\nu}$ and the $SU(3) \times SU(3)$ currents. 

In particular, there was the $U(1)$ problem, which I knew from
Gell-Mann's 1969 Hawaii lectures\cite{hawaii}. If we abstract from the 
free-quark model, the isoscalar current 
\begin{equation}
J^{0}_{\mu 5} = \bar{u}\gamma_\mu\gamma_5 u + \bar{d}\gamma_\mu\gamma_5 d
\end{equation}
is conserved in the $SU(2) \times SU(2)$ limit. This is a disaster\cite{glash}
because then the $SU(2) \times SU(2)$ condensate
\begin{equation}
\langle\mbox{vac}|\bar{u}u + \bar{d}d|\mbox{vac}\rangle \not= 0
\label{cond}
\end{equation}
\emph{also} acts as an axial $U(1)$ condensate. In addition to
$\pi^+, \pi^0, \pi^-$, there would have to be a fourth Nambu-Goldstone 
boson, an isoscalar $0^-$ meson of mass $O(m_\pi)$. If just one extra 
conserved current could cause so much trouble, we certainly did not want 
an infinite tower of them.

The choice between canonical and anomalous dimensions would be cleared 
up by asymptotic freedom\cite{pol,GW} two years later. Only $\theta_{\mu\nu}$ 
and the $SU(N_f) \times SU(N_f)$ currents behave canonically. Coefficients
$\mathcal{C}_n$ of other operators $O_n$ have their canonical
behavior modified by inverse logarithmic powers, corresponding to a 
very weak violation of Bjorken scaling. The $U(1)$ problem was not
so easily dismissed, and while majority opinion is that it is understood,
they all miss the reference\cite{kaiser} where problems yet to be resolved
are analysed.

\section{Approximate Scale Invariance at Low Energies}
\label{three}

At the same time (1969-71), I was supposed to be working on my PhD 
research project. Clearly the hadronic ground state $|\mbox{vac}\rangle$ 
breaks scale invariance very strongly, given the $1$ GeV scale set by 
baryons. This could be simply due to scale invariance being badly broken
explicitly, with the trace $\theta^\mu_\mu$ large as an operator. The 
alternative is that scale invariance is approximately conserved in the 
Nambu-Goldstone mode with a massless $0^{++}$ \emph{dilaton} in the limit 
$\theta^\mu_\mu \to 0$.

The analogy with chiral symmetry was obvious. Both chiral $SU(3) \times SU(3)$
and scale symmetry would be manifest at short distances and hidden elsewhere
by the effects of their Goldstone bosons, the $0^-$ octet $\pi, K, \eta$ and 
the $0^+$ singlet dilaton $\sigma$ (not to be confused with the field 
``$\sigma$'' of the sigma model). I chose the simplest case where scale 
invariance was the result of taking the chiral $SU(3) \times SU(3)$ limit,
so the 3-flavor version of the chiral condensate (\ref{cond}) could also 
act as a scale condensate.

This picture is no longer entirely accurate, given that QCD renormalization 
effects break scale symmetry everywhere, including short distances. However 
if 3-flavor QCD has an infrared fixed point $\alpha_\mathrm{IR}$, where 
$\theta^\mu_\mu$ vanishes apart from $O(m_{u,d,s})$ corrections, the essential 
features of the original scheme can be reproduced by a double expansion in
the running gluon coupling $\alpha_s$ about $\alpha_\mathrm{IR}$ and the
light quark masses $O(m_{u,d,s})$ about zero.\cite{CT}

The dilaton idea is contained in footnote 38 of the 1962 current algebra
paper.\cite{algebra} A ``resonance or quasi-resonance'' which dominates
a dispersion relation for
\begin{equation}
\langle\mbox{particle}|\theta^\mu_\mu|\mbox{particle}\rangle
= \mbox{particle mass}
\label{GTtype} 
\end{equation}
yields ``a relation of the Goldberger-Treiman type'' where ``the coupling
of the resonant state to different particles is roughly proportional to
their masses''. In the scale-invariant limit,\cite{hawaii} the vacuum 
would become degenerate, as for exact chiral symmetry, except for the
degeneracy being noncompact. Physical predictions are then the result 
of expanding in $m_\sigma^2$ about zero. 

The term ``dilaton'' is often used in a manner which is distinct 
from the scheme above or even contradicts it. The earliest variant
was Fujii's proposal\cite{fujii} of a finite-range scalar component 
of gravity. Gell-Mann called it a ``Brans-Dickeon'' after the 
well-known proponents of the scalar-tensor theory of gravity\cite{bd}, 
but the name did not stick. In modern times, ``dilaton'' is often 
used for a scalar particle which has zero mass classically but 
becomes massive due to quantum corrections, such as Higgs bosons 
which acquire mass due to dimensional transmutation.\cite{CW} 
Since there is no way of ``turning off'' such a mass, this has nothing 
to do with dilatons in the original sense. 

In my student days, the main candidate for $\sigma$ was $\epsilon(700)$, 
whose existence was not clear. Final state pions interact very strongly in 
the $0^{++}$ channel, so there was good reason to assume the presence of a 
resonance far off shell. However that meant that it was very hard to pin
down in phase-shift analyses. It was declared dead in the 1976 particle
data tables, but in recent years, has been resurrected as the broad but
clearly defined resonance\cite{Cap} $f_0(500)$. 

If dilatons couple to mass, why is its coupling to pions so large? 
In leading order, one would expect $F_\sigma g_{\sigma\pi\pi}$ to be
$2m_\pi^2$ for the coupling $g_{\sigma\pi\pi} \sigma \boldsymbol{\pi .\pi}$,
where $F_\sigma$ is the analogue of the pion decay constant $F_\pi \simeq 93$ MeV
and has a similar order of magnitude:
\begin{equation} 
\langle\sigma(q)|\theta_{\mu\nu}|\mathrm{vac}\rangle
= (F_\sigma/3)\bigl( q_\mu q_\nu - g_{\mu\nu}q^2\bigr) \,.
\label{F}\end{equation}
The solution, on which I based my PhD thesis, was to note that the
result is really
\begin{equation}
F_\sigma g_{\sigma\pi\pi} = 2m_\pi^2 + O(m_\sigma^2) 
\end{equation}
and use approximate chiral symmetry to deduce the coefficient of 
$m_\sigma^2$:
\begin{equation}
F_\sigma g_{\sigma\pi\pi} = - m_\sigma^2 + O(m_\pi^2) \,.
\label{width}
\end{equation}
This implies a width of a few hundred MeV, as required. I did it 
the hard way, using basic current algebra, and so took too long to
obtain a mass formula for $m_\sigma^2F_\sigma^2$. In the meantime,
John Ellis was working on his PhD in Cambridge (UK), and obtained 
both Eq.~(\ref{width}) and the mass formula by efficient use of a 
chiral-scale effective Lagrangian. A few months later, we met and were
able to compare notes at the 1971 Coral Gables conference.\cite{jelly,rod}

No account of these times would be complete without mentioning the episode
in 1970 when Feynman became excited about Bose statistics for quarks. He
hoped to explain the $\Delta I = 1/2$ rule for nonleptonic decays of strange
particles. How quarks could possibly be bosons was a matter for future study; 
perhaps their bad statistics would not matter if they were confined. Almost 
immediately, we heard that the idea had already been suggested\cite{jap,chls},
but the interest generated by Feynman\cite{FKR} in this key problem was good for 
particle physics. A few months later, the correct version of the idea was 
proposed\cite{PW} (also anticipated in Japan\cite{MM}): for fermion quarks
with color (and even for paraquarks\cite{king}), the color antisymmetrization 
of $qqq$ states plus current algebra implies the $\Delta I = 1/2$ rule for 
nonleptonic hyperon decays, but says nothing about $\Delta I = 1/2$ for 
$K \to \pi\pi$.

Since nonleptonic strange particle decays had been a problem for so 
long,\cite{GMP} my interest was piqued. I told Murray of this, carefully 
avoiding any suggestion that quarks could be bosons (which I didn't believe 
anyway), and drew the response ``watch out, it's a can of worms!'' I was 
too busy finishing my PhD to pursue it; otherwise, I may have drawn Fig.~1, 
which is required by approximate chiral-scale invariance.
\begin{figure}[bt]
\center\includegraphics[scale=.80]{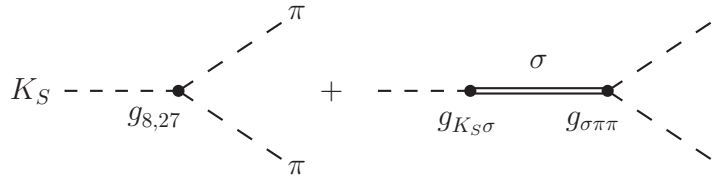}
\caption{Tree diagrams in chiral-scale perturbation theory\cite{CT} for
$K_S\to\pi\pi$. The vertex amplitudes due to \textbf{8} and 
\textbf{27} contact couplings $g_8$ and $g_{27}$ are dominated by the 
$\sigma/f_0$ pole amplitude. The magnitude of $g^{}_{K_S\sigma}$ can be
deduced from $K_S \to \gamma\gamma$ and $\gamma\gamma \to \pi\pi$.}
\label{fig:k_pipi}
\end{figure}%
It shows that the $\Delta I = 1/2$ rule for kaons is due to a large 
contribution from the dilaton pole. Only after 40-odd years, with help 
from my young colleague Lewis Tunstall, can I report a solution to that 
problem.\cite{CT} For hyperon decays, the $\Delta I = 1/2$ rule is 
understood, but current algebra does not seem to work: that part of 
the problem is still a can of worms.

After Coral Gables, there was a thesis to be written, and a suitable way
of ending it had to be found. What else could dilatons do?

From the literature on axial anomalies, I knew about Schwinger's 1951
paper on gauge invariance\cite{julian}. In Sec.~V, he obtained unique
results for both $\pi^0 \to \gamma\gamma$ and $\sigma \to \gamma\gamma$ 
in one-loop Yukawa theory by imposing gauge invariance on the renormalization
procedure. In terms of the electromagnetic field tensor $F_{\mu\nu}$, 
fermion mass $M$, Yukawa coupling $g$, and fine-structure constant $\alpha$,
the answer for $\sigma \to \gamma\gamma$ is
\begin{equation}
{\cal L}^{\mathrm{Yukawa}\rule[-1mm]{0mm}{1.5mm}}_{\rule{0mm}{1.5mm}\sigma\gamma\gamma} 
= - \frac{\alpha g}{6\pi M}\sigma F^{\mu\nu} F_{\mu\nu} 
\end{equation}
In the second-last paragraph of my thesis, I noted that this breaks scale 
invariance (operator dimension $\not= 4$), so if $M$ plays the role of $F_\sigma$ 
as well as $F_\pi$, perhaps both $F_\pi g_{\pi\gamma\gamma}$ and 
$F_\sigma g_{\sigma\gamma\gamma}$ are anomalous. Already, Wilson had shown\cite{kgw2} 
that one-loop corrections in $\lambda\phi^4$ theory break scale invariance, 
which he interpreted as an anomaly in the trace of $\theta_{\mu\nu}$. Perhaps 
there is an electromagnetic trace anomaly due to strong interactions? I was
moving to a post-doctoral job at Cornell; as soon as I arrived, I would try 
to extend Wilson's method for $\pi^0 \to \gamma\gamma$ to $\sigma \to 
\gamma\gamma$.

\section{Derivation of $\pi^0 \to \gamma\gamma$ for Nonperturbative Pions}

When Schwinger analysed $\pi^0 \to \gamma\gamma$, chiral invariance and PCAC 
(partially conserved axial current) were unknown. At issue was the equivalence 
\begin{equation}
\phi\bar{\psi}\gamma_5\psi\ \leftrightarrow\ 
  - \bigl(i/2M\bigr)\bar{\psi}\gamma_\mu\gamma_5\psi \del^\mu\phi \,,
\quad \phi = \pi^0 \mbox{ field}
\end{equation}
between pseudoscalar and pseudovector couplings for the one-fermion-loop 
triangle diagram. The trouble was that the product of the fermion fields at 
the same point is singular. The solution was to consider $\psi$ and 
$\bar{\psi}$ at different points $x'$ and $x''$ and make the analysis 
gauge invariant: then the limit $x' \to x''$ becomes finite. Rephrased
in terms of chiral symmetry, the problem is that the Noether construction
fails because (a) it requires 
\begin{equation}
\frac{\del\mathcal{L}}{\del\del^\mu\psi} = \bar{\psi}\gamma_\mu\ \mbox{ and }\ 
\delta^{}_{\mathrm{axial}}\psi = \gamma_5\psi 
\end{equation}
to be multiplied at the same point and (b) it does not work for non-local
expressions produced by point splitting.  The axial anomaly\cite{bell,steve} 
is responsible for this failure: it is the finite counterterm mismatch between 
gauge invariant and chiral invariant renormalization prescriptions for 
axial-vector operators.

Wilson's version of this was designed to avoid perturbation theory. In 
particular, if pions are $q\bar{q}$ states which become Nambu-Goldstone bosons 
in the chiral limit, they are certainly not perturbative and so should not 
be represented by a perturbative field $\phi$. 

The other key feature of his approach was the use of short distance analysis. 
The connection between axial and trace anomalies and short distance behavior 
is best illustrated by considering first how equal-time commutators produce 
contact terms $\sim \delta^4(x-y)$ in ordinary Ward identities. 

Given a free massive boson field $\varphi$, let $\del_\mu\varphi$ play the 
role of a current. Canonically, the divergence of $T\{\varphi\del_\mu\varphi\}$ 
is found by writing the $T$-product in terms of step functions 
$\theta(\pm x_0)$ and unordered field products, differentiating the 
step functions
\begin{equation}
\frac{\del\ }{\del x^\mu}\theta(\pm x_0) = \pm\delta(x_0)g_{0\mu}
\end{equation}
and substituting $\del^2\varphi = - m^2\varphi$:
\begin{equation}
\del^\mu T\bigl\{\varphi(0)\del_\mu\varphi(x)\bigr\} 
= \bigl[\del_0\varphi(x),\varphi(0)\bigr]\delta(x_0) 
 - m^2T\bigl\{\varphi(0)\varphi(x)\bigr\} \,.
\label{canon1}
\end{equation}
In this case, the contact term can be found by substituting a canonical 
commutator:
\begin{equation}
\bigl[\del_0\varphi(x),\varphi(0)\bigr]\delta(x_0)
= - i\delta^4(x)I\,, \quad I = \mbox{identity operator.}
\label{canon2}
\end{equation}
The short-distance method is to note that a term $\sim \delta^4(x)$ can 
arise only if $\del^\mu$ acts on a singularity $\sim 1/x^3$ at $x \sim 0$. 
The leading term of the operator product expansion for 
$T\{\varphi(0)\del_\mu\varphi(x)\}$ is given by the propagator of the 
massless theory
\begin{equation}
T\{\varphi(0)\del_\mu\varphi(x)\} \to \frac{x_\mu}{2\pi^2(x^2 - i\epsilon)^2}I\,,
\quad I = \mbox{identity operator}
\end{equation}
Substituting $\del^\mu\bigl(x_\mu/x^4\bigr) = -2i\pi^2\delta^4(x)$ and  
$\del^2\varphi = - m^2\varphi$, we find
\begin{equation}
\del^\mu T\bigl\{\varphi(0)\del_\mu\varphi(x)\bigr\} 
= - i\delta^4(x)I
 - m^2T\bigl\{\varphi(0)\varphi(x)\bigr\} \,.
\end{equation}
in agreement with Eqs.~(\ref{canon1}) and (\ref{canon2}).

For the axial anomaly, the problem is to evaluate the quantity
\begin{equation}
S = -\frac{\pi^2}{12}\epsilon^{\mu\nu\alpha\beta}\iint d^4x d^4y\, x_\mu y_\nu
 T\langle\mbox{vac}|J_\alpha(x)J_\beta(0)\del^\gamma J_{\gamma 5}(y)|\mbox{vac}\rangle
\,,
\label{S}
\end{equation}
where data for $\pi^0 \to \gamma\gamma$  and approximate $SU(2) \times SU(2)$ 
symmetry imply $S \simeq + 0.5$.  The constant $S$ normalizes the contact term 
in an anomalous Ward identity of the form
\begin{align}
&\del_y^\nu\mbox{``$T$''}
\langle\mbox{vac}|J_\alpha(x)J_\beta(0)J_{\nu 5}(y)|\mbox{vac}\rangle
\nonumber \\
&= 
\frac{S}{2\pi^2}\epsilon_{\alpha\beta\mu\nu}\del_x^\mu\del_y^\nu\delta^4(x)\delta^4(y)
 + T\langle\mbox{vac}|J_\alpha(x)J_\beta(0)\del^\nu J_{\nu 5}(y)|\mbox{vac}\rangle 
\,.
\label{contact}
\end{align}
The contact term scales as $1/\{\mbox{length}\}^{10}$, so it must be generated
by a short distance singularity 
\begin{equation} 
J_\alpha(x)J_\beta(0)J_{\nu 5}(y) \sim 1/\{\mbox{length}\}^9
\end{equation}
as \emph{both} $x_\mu$ and $y_\mu$ tend to zero. (Do not confuse this with
the short-distance properties of $J_\alpha J_\beta \del^\gamma J_{\gamma 5}$ in 
Eq.~(\ref{S}), where the condition dim\ $\del^\gamma J_{\gamma 5} < 4$ ensures 
convergence of the integral.)

In Eq.~(\ref{contact}), a \emph{single} derivative $\del_y^\mu$ produces 
\emph{a product of two delta functions}, so it is clear that $\theta$-functions 
in time \emph{cannot} be used to construct ``$T$''. This example exposes 
the limitations of canonical field theory very effectively.

In perturbation theory, it has long been known\cite{SP} but not often noted 
that time ordering is part of the renormalization procedure. In general,
``$T$'' must be regarded as an operation which depends on the 
renormalization prescription. The difference between two time-ordering
procedures for a given operator product is a set of contact terms at
coinciding points. In the case of the triangle diagram coupled to photons,
electromagnetic gauge invariance specifies the renormalization procedure
completely:
\begin{equation} 
\mbox{``$T$''}  \rightarrow  T_{\mathrm{e'mag}}  \,.
\end{equation}

Wilson\cite{kgw} circumvented the ``$T$'' problem by excising a small 
neighbourhood around lines of coinciding points in the integral (\ref{S}).
Let the region of integration be restricted to the region 
\begin{equation}
\mathcal{R} = \bigl\{|x_0| > \epsilon\,,\ |y_0| > \epsilon'\,,\ 
|x_0 - y_0| > \epsilon''\bigr\} \,.
\label{region}
\end{equation}
\begin{figure}[htb]
\center\includegraphics[scale=.95]{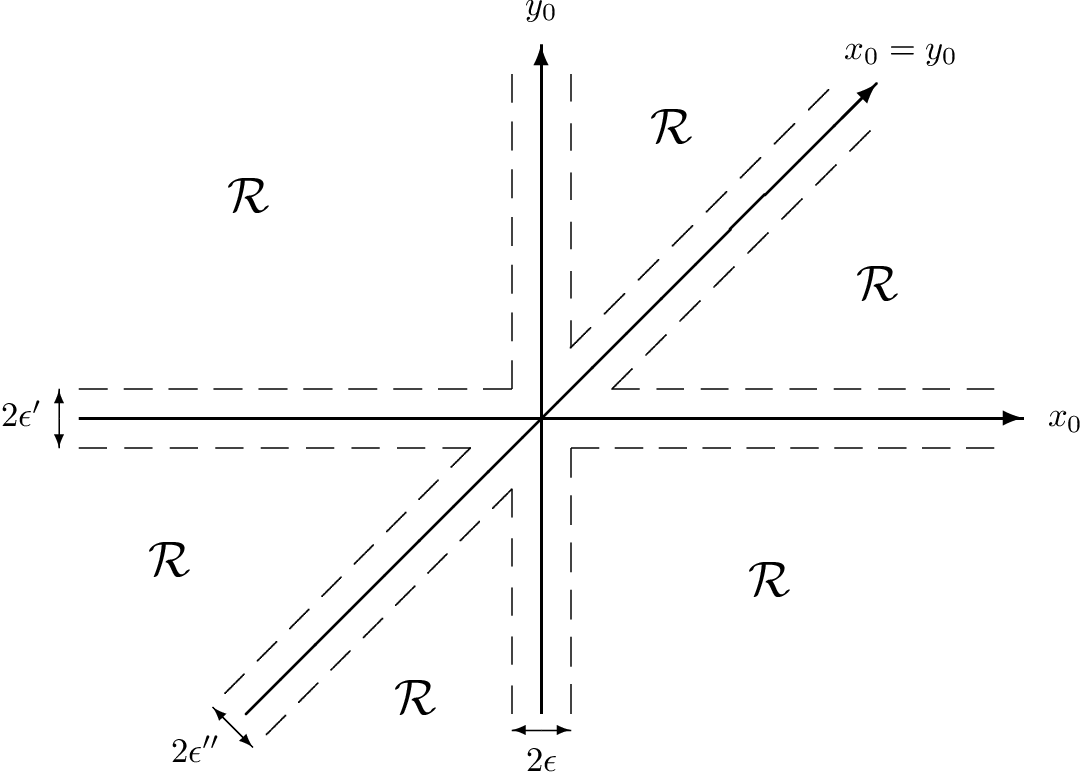}
\caption{Integration region $\mathcal{R}$ defined by Eq.~(\ref{region}).
\label{excise}}
\end{figure}
shown in Fig.~\ref{excise}, so that Eq.~(\ref{S}) becomes
\begin{equation}
S = - \frac{\pi^2}{12}\epsilon^{\mu\nu\alpha\beta}\iint\limits_\mathcal{R} 
 d^4x d^4y\, x_\mu y_\nu
 T\langle\mbox{vac}|J_\alpha(x)J_\beta(0)\del^\gamma J_{\gamma 5}(y)|\mbox{vac}\rangle
 + O(\epsilon,\epsilon', \epsilon'') \,.
\label{S'}
\end{equation}
Of course, $S$ does not depend on $\epsilon$, $\epsilon'$, or $\epsilon''$.
Within $\mathcal{R}$, define
\begin{align}
X_\gamma &= \epsilon^{\mu\nu\alpha\beta}x_\mu y_\nu T\langle\mbox{vac}|
  J_\alpha(x)J_\gamma(0)J_{\beta 5}(y) + J_\gamma(x)J_\alpha(0)J_{\beta 5}(y)
  |\mbox{vac}\rangle  \nonumber \,,\\
Y_\gamma &= \epsilon^{\mu\nu\alpha\beta}x_\mu y_\nu T\langle\mbox{vac}|
  J_\alpha(x)J_\beta(0)J_{\gamma 5}(y) + J_\alpha(x)J_\gamma(0)J_{\beta 5}(y)
  |\mbox{vac}\rangle
\end{align}
where now time ordering with $\theta$-functions is allowed because  
$\mathcal{R}$ excludes coinciding points. This also means that derivatives
commute with the $T$-operation, so we can obtain
\begin{equation}
S  = - \frac{\pi^2}{12}\iint\limits_\mathcal{R} 
 d^4x d^4y\,\bigl(\del^\gamma_x X_\gamma + \del^\gamma_y Y_\gamma\bigr)
  + O(\epsilon,\epsilon', \epsilon'')
\end{equation}
by using current conservation $\del^\gamma J_\gamma = 0$, translation invariance
of $|\mbox{vac}\rangle$ and symmetry $x \leftrightarrow y$ of the integral
to $O(\epsilon,\epsilon', \epsilon'')$. If $\Sigma$ is the surface in
8-dimensional space which bounds $\mathcal{R}$, we have
\begin{equation}
S  = - \frac{\pi^2}{12}\int_\Sigma\!d\vec{\Sigma}\cdot\vec{Z} 
  + O(\epsilon,\epsilon', \epsilon'')
\label{surface}
\end{equation}
where $\vec{Z} = \bigl(X_\gamma , Y_\gamma\bigr)$ is an 8-dimensional vector
formed from the components of $X_\gamma$ and $Y_\gamma$. 

So $S$ is given by the result of taking $\epsilon$, $\epsilon'$ and 
$\epsilon''$ to zero in Eq.~(\ref{surface}). Since the current operators 
commute at space-like 
separations, their products at short distances are all that we need.
If we consider (say) $\epsilon \to 0$ and exclude the $x,y \sim 0$ 
neighbourhood where the axes in Fig.~\ref{excise} meet, we have 
$x \sim 0$ for fixed $y$, which means expanding in $J_\alpha(x)J_\beta(0)$ 
to produce an equal-time commutator. There could be three commutators in 
principle, one for each axis, but explicit checks confirm the 
conclusion\cite{suth} that they all vanish. Therefore $S$ is entirely
determined by the leading VVA short-distance singularity 
\begin{equation}
T\{J_\alpha(x)J_\beta(0)J_{\gamma 5}(y)\} \sim G_{\alpha\beta\gamma}(x,y)I \,,
\quad x,y \sim 0 \,,
\label{VVA}
\end{equation}
so it can be calculated if the three-point function $G_{\alpha\beta\gamma}$ 
is known.

At this point, I tried the same analysis for the trace anomaly.
Let the amplitude for photons to couple to the hadronic 
energy-momentum tensor be
\begin{equation}
\langle\gamma(\epsilon_1,k_1)\gamma(\epsilon_2,k_2)|
        \theta^\mu_\mu(0)|\mbox{vac}\rangle
= \bigl(\epsilon_1\cdot\epsilon_2 k_1\cdot k_2 
        - \epsilon_1\cdot k_2 \epsilon_2\cdot k_1\bigr)F\bigl((k_1 + k_2)^2\bigr)
\,.
\label{ampl}
\end{equation}
As in Eq.~(\ref{S}) for $S$, the trace anomaly corresponds to the low-energy 
limit $k_1, k_2 \sim 0$:
\begin{equation}
F(0) = - \frac{\pi\alpha}{3}\iint d^4xd^4y\,x\cdot y
       \langle\mbox{vac}|J^\alpha(x)J_\alpha(0)\theta^\mu_\mu(y)|\mbox{vac}\rangle
\,. \label{trace}
\end{equation}
The aim was to substitute the formula for the divergence of the
conformal current
\begin{equation}
 \del^\mu_y\bigl\{(2y_\lambda y^\nu - \delta^\nu_\lambda y^2)\theta_{\mu\nu}(y)\bigr\}
 = 2y_\lambda\theta^\mu_\mu(y)
\label{conf}
\end{equation}
and integrate by parts. To my surprise, I found that it was not necessary to
exclude coinciding points as in Fig.~\ref{excise}. Instead, I found that an 
answer could be found directly by restricting just the $x$ integration
to $|x_0| > \eta$ for small $\eta > 0$ to keep the $x,y \sim 0$ 
singularity
\begin{equation}
T\{J_\alpha(x)J_\beta(0)\theta_{\mu\nu}(y)\} \sim K_{\alpha\beta\mu\nu}(x,y)I
\end{equation}
under control. Then integration by parts with respect to $y$ produced
known equal-time commutators, so the $y$ integral could be done, with the
result
\begin{equation}
F(0) = -\frac{i\pi\alpha}{6}\int_{|x_0| > \eta}\!d^4x\,\del^\nu_x\{x^2 x_\nu
      T\langle\mbox{vac}|J^\alpha(x)J_\alpha(0)|\mbox{vac}\rangle\} + O(\eta)\,.
\end{equation}
We have $J_\alpha J_\beta \sim R/x^6$ at short distances, where $R$ is the 
asymptotic Drell-Yan ratio
\begin{equation}
R = \bigl\{(\sigma(e^+e^- \to \mbox{hadrons})
    \bigl/(\sigma(e^+e^- \to \mu^+\mu^-)\bigr\}_{\mathrm{energy} \to \infty}
\end{equation}
so the $x$ integral can also be done, yielding an exact result:
\begin{equation}
F(0) = 2R\alpha/3\pi  \,.
\label{trace2}
\end{equation}
In effect, an anomalous term%
\footnote{The extrapolation in $(k_1 + k_2)^2$ from zero to 
$m_\sigma^2$ used to estimate $F_\sigma g_{\sigma\gamma\gamma}$ from $F(0)$
has had to be modified,\cite{CT} because $\pi,K$ loop diagrams compete 
with the $\sigma$-pole amplitude.}
$(R\alpha/6\pi)F_{\mu\nu}F^{\mu\nu}$ is induced in the trace of the 
energy-momentum tensor by electromagnetism.\cite{rjc}
The same result was found independently by Mike Chanowitz and 
John Ellis\cite{CE} via a momentum-space analysis. It was the 
immediate precursor of the gluonic trace anomaly 
$\beta(\alpha_s)/(4\alpha_s) G^a_{\mu\nu}G^{a\mu\nu}$
found a few years later.\cite{mink}.

The unexpected feature of the analysis leading to Eq.~(\ref{trace2}) was 
that, although the 3-point singularity $ K_{\alpha\beta\mu\nu}$ is responsible 
for the presence of the trace anomaly, its full functional form is not 
needed: only the subregion $x-y \ll x,y$ \emph{within} the $x,y \sim 0$ 
region is needed. In Fig.~\ref{excise}, this subregion connects the central
area $x,y \sim 0$ to other $x \sim y$ regions along the diagonal
axis $x_0 = y_0$.

That led me to consider \emph{nested} operator product expansions, where
an expansion such as 
\begin{equation}
T\{A(x)B(0)\} \sim \sum_m \mathcal{C}_m(x) O'_m(0)\ \mbox{ for } x \sim 0
\end{equation}
is substituted into a larger expansion, e.g.
\begin{equation}
T\{A(x)B(0)C(y)\} \sim \sum_n f_n(x,y) O_n(0)\ \mbox{ for } \quad x,y \sim 0\,.
\end{equation}
This is legitimate provided that $y$ is independent of the limit $x \to 0$,
i.e.\ $x \ll y$. Then a subsequent limit $y \to 0$ can be taken:
\begin{equation}
T\{O'_m(0)C(y)\} \sim \sum_n \mathcal{C}_{mn}(y)O_n(0) \,.
\end{equation}
The result is a set of consistency conditions\cite{rjc}
\begin{equation}
 f_n(x,y) \sim \sum_m \mathcal{C}_m(x)\mathcal{C}_{mn}(y)  \,.
\end{equation}
The idea works at short distances (and not on other parts of light cones)
provided the limits are \emph{nested}. For the example above, fix $\hat{x}$ 
and $\hat{y}$ in
\begin{equation}
x = \rho_1\rho_2 \hat{x}\ \mbox{ and }\ y = \rho_2 \hat{y}
\end{equation}
and take the limits $\rho_1 \to 0$ and $\rho_2 \to 0$ independently. 
This is the position-space version of Weinberg's limiting 
procedure\cite{wein} used to classify the asymptotic behavior of 
amplitudes and hence justify power counting methods for renomalization.
 
An obvious next step was to apply this procedure to the short-distance 
VVA function $G_{\alpha\beta\gamma}$ of Eq.~(\ref{VVA}). Let 
\begin{equation}
u = x^2 -i\epsilon\ , \quad v =  y^2 -i\epsilon\ , \quad w = (x-y)^2 -i\epsilon \,.
\end{equation}
The relevant two-point expansions are
\begin{align}
T\{J_\alpha(x)J_\beta(0)\} 
&\sim R(g_{\alpha\beta}x^2 - 2x_\alpha x_\beta)I\bigl/(\pi u)^4 
 + K\epsilon_{\alpha\beta\lambda\mu}x^\lambda J^\mu_{\ 5}(0)\bigl/(3\pi^2u^2)
\nonumber \\
T\{J^\mu_{\ 5}(0)J_{\gamma 5}(y)\} 
&\sim R'(\delta^\mu_\gamma y^2 - 2y^\mu y_\gamma)I\bigl/(\pi v)^4
\end{align}
where $R'$ is the isovector part of $R$, and $K$ is measurable in polarised
deep-inelastic electroproduction or in $e^+ + e^- \to \mu^+ + \mu^- + \pi^0$.
The result
\begin{equation}
G_{\alpha\beta\gamma}(x,y) \underset{x \ll y}{\longrightarrow}
\bigl\{K\epsilon_{\alpha\beta\lambda\mu}x^\lambda J^\mu_{\ 5}(0)\bigl/(3\pi^2u^2)\bigr\}
 R'(\delta^\mu_\gamma y^2 - 2y^\mu y_\gamma)I\bigl/(\pi v)^4  \,.
\end{equation}
said something about the normalization of $G_{\alpha\beta\gamma}$, but without
a formula valid for the whole $x, y \sim 0$ region, the calculation of $S$ 
could not be completed.

In the meantime, I was checking products of $\theta_{\mu\nu}$ with other
currents to see if the absence of the soft trace at short distances would 
imply asymptotic%
\footnote{This has \emph{nothing} to do with the properties of the vacuum 
state. As noted at the beginning of Sec.~\ref{three}, $|\mbox{vac}\rangle$ 
breaks scale and hence conformal invariance very strongly. This may be
due to explicit symmetry breaking or to the symmetry being realised in the 
Nambu-Goldstone mode.}
conformal invariance, as indicated by Eq.~(\ref{conf}).
Satisfied that it did, I required conformal invariance for $G_{\alpha\beta\gamma}$,
found that it had to be proportional to the triangle diagram, and then
found that this result had already been published by Schreier.\cite{ethan}

So the evaluation of the VVA singular function was complete:
\begin{equation}
G_{\alpha\beta\gamma}(x,y) 
= \frac{KR'}{12\pi^6u^2 v^2 w^2}\mbox{Tr}\bigl\{\gamma_\alpha\gamma\cdot x
 \gamma_\beta\gamma\cdot y\gamma_\gamma(\gamma\cdot x - \gamma\cdot y)
  \gamma_5\bigr\}  \,.
\end{equation}
When $G_{\alpha\beta\gamma}$ is substituted into Eq.~(\ref{surface}), the
equal-time commutator regions give no contribution (as before), so as long 
as (say) $\epsilon'$ is held fixed, the limits $\epsilon \to 0$ and 
$\epsilon'' \to 0$ can be taken without intruding on the short-distance 
region. Thus%
\footnote{These details, taken from a letter I wrote to Fritzsch and Gell-Mann 
at the time,\cite{cao} should have been part of ref.~[10] of my paper\cite{rjc}
but it was never finished.}
\begin{equation}
S = \frac{\pi^2}{12}\int\!d^3y\int\!d^4x\,
  \bigl\{\widetilde{Y}_0(y_0=\epsilon')-\widetilde{Y}_0(y_0=-\epsilon')\bigr\}
 + O(\epsilon')
\end{equation}
where $\widetilde{Y}_\gamma$ is the $x,y \sim 0$ part of $Y_\gamma$:
\begin{align}
\widetilde{Y}_\gamma &= \epsilon^{\mu\nu\alpha\beta}x_\mu y_\nu
  \bigl\{G_{\alpha\beta\gamma}(x,y) + G_{\alpha\gamma\beta}(x,y)\bigr\} 
\nonumber \\
 &= - \frac{4KR'}{3\pi^6 u^2 v^2 w^2}y_\gamma\bigl\{x^2 y^2 - (x\cdot y)^2\bigr\} \,.
\end{align}
Do the $x$-integral
\begin{equation}
\int\!d^4x\,\{x^2 y^2 - (x\cdot y)^2\}/(uw)^2 = - 3\pi^2i/2
\end{equation}
and then the $y$-integral
\begin{equation}
\int\!d^3y/v^2 = - \pi^2i/|y_0|
\end{equation}
to obtain the desired formula\cite{rjc}
\begin{equation}
3S = KR'  \,.
\end{equation}
It relates the low-energy amplitude $S$ to high-energy amplitudes $R'$ 
and $K$. 

Being anxious to avoid model dependence, I allowed for the possibility
that the electromagnetic current is not a pure $SU(3)$ octet,
\begin{equation}
4R' \leqslant 3R \,.
\end{equation}
At that time, we did not know $R$, $R'$ or $K$. In particular, data showing 
scaling behavior for $e^+ + e^- \to$ hadrons was not available until the 1974
London Conference, a year after asymptotic freedom. The value of $S$ could 
well be exactly 0.5, so Adler's non-renomalization theorem\cite{steve,ab} for 
$S$ suggested a theory with three species of quark, but I could not see how 
to deal with colored electromagnetic currents or paraquark operators.

So when I received by return mail a letter from Gell-Mann proposing colored 
fractional quarks with color-neutral currents, it seemed to me that this 
clarified matters from the point of view of Adler's theorem, but I felt 
(for reasons discussed above) that having free quarks on the light cone 
was going too far. However, sometimes an oversimplification can lead to 
correct answers --- in this case, QCD\cite{fgm2} and asymptotic 
freedom.\cite{pol,GW}

Initially, the QCD proposal looked good as a model of constituent quarks, 
but not for current quarks and the $\pi^0 \to \gamma\gamma$ analysis. 
Having found the electromagnetic trace anomaly, we knew already that 
$\theta^\mu_\mu$ would have anomalous gluonic terms proportional to 
$G^a_{\mu\nu}G^{a\mu\nu}$ which would break scale and conformal invariance
at short distances.

What asymptotic freedom did was to turn QCD into a good theory of current 
quarks as well as constituent quarks. The breaking of scale invariance at
short distances was minimal, being associated with operators which
are not conserved exactly or partially. The analysis of $\pi^0 \to \gamma\gamma$ 
can be still be carried through since all of the equations remain valid: 
they can be derived by using asymptotic freedom instead of asymptotic 
conformal invariance. The results for three colors are
\begin{equation}
R = 2,\ R' = 1.5,\ K = 1, \mbox{ and } S = 0.5
\end{equation}
where the non-renormalization theorem for $S$ is \emph{not} used.
All of this goes through \emph{without} treating pions perturbatively. 

The method of nested operator product expansions is now not needed for 
the $\pi^0 \to \gamma\gamma$ derivation, but it is generally valid in 
renormalized field theory. As a result, coupling constant dependence
of the form
\begin{equation}
3S = K(g) R'(g)
\end{equation}
can be investigated.\cite{ACGJ} This program has been extensively pursued 
by Andrei Kataev, Stan Brodsky and their collaborators.\cite{andrei}

In retrospect, there came a time when abstracting physics had to give way
to guessing the correct model. I remember that time well.

%\bibliographystyle{ws-rv-van}
%\bibliography{ws-rv-sample}
%\blankpage
%\printindex[aindx]                 % to print author index
%\printindex                         % to print subject index
\end{document}